\documentclass{aa}

\usepackage{graphicx}
\usepackage{color}

\def\degr{\mathrm{^\circ}}
\def\micron{\,\mu\mathrm{m}}

\def\rsun{\,\mathrm{R}_{\sun}}

\def\aap{A\&A}
\def\aaps{A\&AS}
\def\apj{ApJ}

\def\mnras{MNRAS}

\begin{document}


\title{VLTI/AMBER and VLTI/MIDI spectro-interferometric \\ observations of the B[e] supergiant
CPD$-57\degr\,2874$\thanks{Based on observations obtained with the
VLTI of the European Southern Observatory (ESO), Chile}}
%
%
\subtitle{Size and geometry of the circumstellar envelope in the near- and mid-IR}
\authorrunning{Domiciano~de~Souza et al.}
\titlerunning{VLTI/AMBER and VLTI/MIDI observations of the sgB[e] CPD$-57\degr\,2874$}
\author{A.~Domiciano~de~Souza \inst{1}
\and  T.~Driebe\inst{1}
\and  O.~Chesneau\inst{2}
\and  K.-H.~Hofmann\inst{1}
\and  S.~Kraus \inst{1}
\and \\ A.~S.~Miroshnichenko \inst{1,3}
\and  K.~Ohnaka \inst{1}
\and R.~G.~Petrov \inst{4}
\and  Th.~Preibisch\inst{1}
\and P.~Stee \inst{2}
\and \\ G.~Weigelt \inst{1}
\and F.~Lisi \inst{5}
\and F.~Malbet \inst{6}
%
%
\and A.~Richichi \inst{7}
}

\institute{Max-Planck-Institut f\"{u}r Radioastronomie, Auf dem
H\"{u}gel 69, 53121 Bonn, Germany \and Observatoire de la C{\^o}te
d'Azur, Gemini, CNRS UMR 6203, Avenue Copernic, 06130 Grasse,
France \and Dept. of Physics and Astronomy, P.O. Box 26170,
University of North Carolina at Greensboro, Greensboro, NC
27402--6170, USA \and Laboratoire Universitaire d'Astrophysique de
Nice (LUAN), CNRS UMR 6525, UNSA, Parc Valrose, 06108 Nice, France
\and INAF-Osservatorio Astrofisico di Arcetri, Istituto Nazionale
di Astrofisica, Largo E. Fermi 5, I-50125 Firenze, Italy \and
Laboratoire d'Astrophysique de Grenoble, UMR 5571 Universit{\'e}
Joseph Fourier/CNRS, BP 53, 38041 Grenoble Cedex 9, France \and
European Southern Observatory, Karl Schwarzschild Strasse 2, 85748
Garching, Germany }

\offprints{A.~Domiciano~de~Souza,\\\email{Armando.Domiciano@unice.fr}}

\date{Received $<date>$ / Accepted $<date>$}

\abstract{We present the first high spatial and spectral
observations of the circumstellar envelope (CSE) of a B[e]
supergiant (CPD$-57\degr\,2874$), performed with the Very Large
Telescope Interferometer (VLTI). Spectra, visibilities, and
closure phase, were obtained using the beam-combiner instruments
AMBER (near-IR interferometry with three 8.3~m Unit Telescopes or
UTs) and MIDI (mid-IR interferometry with two UTs). The
interferometric observations of the CSE are well fitted by an
elliptical Gaussian model with FWHM diameters varying linearly
with wavelength. Typical diameters measured are
$\simeq1.8\times3.4$~mas or $\simeq4.5\times8.5$~AU (adopting a
distance of $2.5$~kpc) at $2.2\micron$, and $\simeq12\times15$~mas
or $\simeq30\times38$~AU at $12\micron$. The size of the region
emitting the Br$\gamma$ flux is $\simeq2.8\times5.2$~mas or
$\simeq7.0\times13.0$~AU. The major-axis position angle of the
elongated CSE in the mid-IR ($\simeq144\degr$) agrees well with
previous polarimetric data, hinting that the hot-dust emission
originates in a disk-like structure. In addition to the
interferometric observations we also present new optical
($UBVR_{c}I_{c}$) and near-IR ($JHKL$) broadband photometric
observations of CPD$-57\degr\,2874$. Our spectro-interferometric
VLTI observations and data analysis support the non-spherical CSE
paradigm for B[e] supergiants.  }

\maketitle

\section{Introduction}

Supergiant B[e] (sgB[e]) stars are luminous ($\log L/L_{\sun} >
4.0$) post-main sequence objects showing the B[e] phenomenon
(Lamers et al. 1998): (1) strong Balmer emission lines, (2)
low-excitation emission lines of Fe {\sc ii}, [Fe {\sc ii}], and
[O {\sc i}], and (3) strong near/mid-infrared (IR) excess due to
hot circumstellar dust. Spectroscopic and polarimetric
observations suggest that sgB[e] stars have non-spherical
circumstellar envelopes (CSE; e.g., Zickgraf et al. 1985;
Magalh{\~a}es 1992). Zickgraf et al. (1985) proposed an empirical
model of the sgB[e] CSE that consists of a hot and fast
line-driven wind in the polar regions, and a slow, much cooler and
denser wind (by a factor of $10^2-10^3$) in the equatorial region,
where dust could be formed. Rapid rotation of the central star
seems to play a key role in the origin of the CSE, but a complete
explanation of its formation mechanism is still unknown.

To investigate these crucial questions concerning the origin,
geometry, and physical structure of the sgB[e] CSE, it is
necessary to combine several observing techniques. In particular,
the high spatial resolution provided by optical/IR long-baseline
interferometry allows us to directly probe the vicinity of these
complex objects. In this paper we present the first direct
multi-wavelength measurements of the close environment of a
Galactic sgB[e] star, namely CPD$-57\degr\,2874$ (WRAY 15-535),
using the VLTI with its instruments AMBER and MIDI.

CPD$-57\degr\,2874$ is a poorly-studied object, for which McGregor
et al. (1988) suggested a distance of $d=2.5$ kpc, assuming that
it belongs to the Carina OB association. A high reddening and the
presence of CO emission bands at $2.3-2.4 \micron$ makes it
compatible with the sgB[e] class. Zickgraf (2003) obtained
high-resolution optical spectra exhibiting double-peaked emission
lines that are suggestive of a flattened CSE geometry, typical for
sgB[e] stars. However, the physical parameters of neither the star
nor its CSE have been studied in detail yet.


\section{Interferometric observations and data reduction}
\label{observations}

\subsection{VLTI/AMBER (near-IR)}

CPD$-57\degr\,2874$ was observed on 2005 February 25 using the
VLTI/AMBER instrument (e.g., Petrov et al. 2003) to combine the
light from the 8.3~m Unit Telescopes UT2, UT3, and UT4. With an
exposure time of 85~ms, 3000 spectrally dispersed interferograms
(frames) were recorded on the target and calibrator (HD~90393).
This allowed us to obtain spectra as well as wavelength-dependent
visibilities and a closure phase in the $K$ band with a spectral
resolution of $R = 1500$ between $2.09$ and $2.24\micron$
(including the Br$\gamma$ line).

Data reduction was performed with the \textit{amdlib} software
(Millour et al. 2004; Malbet et al. 2005). We checked the
consistency of our results by selecting a fixed percentage of
frames from the target and calibrator data sets, based on the
fringe contrast signal-to-noise ratio. By keeping 50\%, 30\%, and
10\% of the frames with the best SNR, we found that the derived
quantities were stable (differences $\la4\%$). Moreover, we also
found good agreement between the results from the \textit{amdlib}
software and our own software based on a power spectrum analysis.

\subsection{VLTI/MIDI (mid-IR)}

We also observed CPD$-57\degr\,2874$ with the VLTI/MIDI instrument
(Leinert et al. 2004) on 2004 December 28 and 30. The $N$--band
spectrum as well as spectrally dispersed fringes have been
recorded between $7.9$ and $13.5\micron$ with a spectral
resolution of $R = 30$, allowing us to study the wavelength
dependence of the apparent size of CPD$-57\degr\,2874$ in the
mid-IR. In total, 4 data sets have been obtained using the
UT2-UT3-47\,m and UT3-UT4-62\,m baselines. Several calibrator
stars were observed: HD~37160, HD~50778, HD~94510, and HD~107446.

Data reduction was performed with the MIA (Leinert et al. 2004)
and EWS (\cite{jaffe04}) packages. While MIA follows the classical
power spectrum analysis, in the EWS software the fringes are
coherently added after correction for the instrumental and
atmospheric delay in each scan. The visibilities derived with both
softwares agree within the uncertainties of $\simeq10\%$.

\smallskip

\begin{figure}[t]
 \centering
  \includegraphics*[width=5.5cm,draft=false]{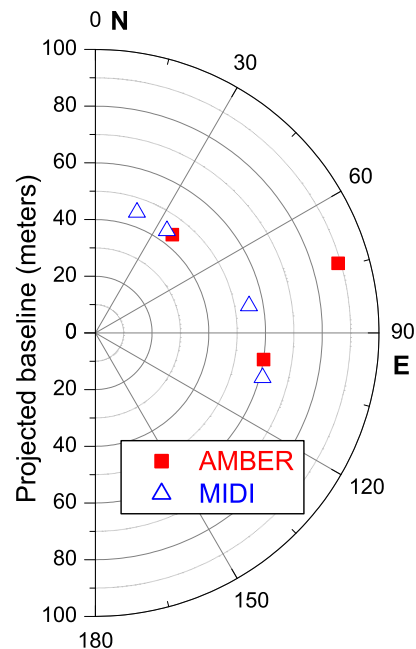}
  \caption{Projected baselines and corresponding position
angles for the AMBER and MIDI observations of
CPD$-57\degr\,2874$ (see also Table~\ref{ta:observation_log}).}
  \label{fig:Bproj_PA}
%
\end{figure}

{The logs of the AMBER and MIDI observations are given in
Table~\ref{ta:observation_log}, while Fig.~\ref{fig:Bproj_PA}
shows the projected baseline lenghts $B_\mathrm{p}$ and
corresponding position angles PA used. In
Table~\ref{ta:observation_log_calib} we list the uniform disc
diameters $\theta_{\mathrm{UD}}$ and observation log for the
calibrators. Calibrated visibilities from both AMBER and MIDI
observations were obtained using the known uniform disk diameters
of the calibrator stars (Richichi et al.~2005), which were
observed in the same nights as CPD$-57\degr\,2874$.}

%

%
\begin{table}
\caption[]{AMBER and MIDI observation log for
CPD$-57\degr\,2874$.} \label{ta:observation_log}
\begin{center}
\begin{tabular}{c}
\textbf{AMBER}  ($2.09\leq \lambda \leq 2.24\micron$) \\
3 Unit Telescopes
\end{tabular}
\begin{tabular}{c c c c c}\hline
night      & $t_{\rm obs}$  &  UT         & $B_\mathrm{p}$ & PA \\
           &   (UTC)        & baseline    &    (m)        &  ($\degr$)   \\ \hline
2005-02-26 & 03:41:21            & UT2-UT3 & 43.5 & 37.8 \\
           &                     & UT3-UT4 & 59.8 & 98.8 \\
           &                     & UT2-UT4 & 89.4 & 73.6 \\
\hline
\end{tabular}
\begin{tabular}{c}
\\
\textbf{MIDI}  ($7.9\leq \lambda \leq 13.5\micron$) \\
2 Unit Telescopes
\end{tabular}
\begin{tabular}{c c c c c}\hline
night      & $t_{\rm obs}$  &  UT         & $B_\mathrm{p}$ & PA \\
           &   (UTC)        & baseline    &    (m)        &  ($\degr$)   \\ \hline
2004-12-29 & 05:52:12            & UT2-UT3  & 45.2           & 18.5        \\
           & 07:26:06            & UT2-UT3  & 43.9           & 35.1        \\ \hline
2004-12-31 & 06:04:03            & UT3-UT4  & 54.8           & 79.6        \\
           & 08:02:48            & UT3-UT4  & 60.9           &104.8        \\ \hline
\end{tabular}
\end{center}
\end{table}
\begin{table}
\caption[]{Uniform disc diameters and observation log for the
calibrators.} \label{ta:observation_log_calib}
\begin{center}
\begin{tabular}{c}
\textbf{AMBER}  ($2.09\leq \lambda \leq 2.24\micron$) \\
3 Unit Telescopes
\end{tabular}
\begin{tabular}{c c c c}
\hline
Calibrator & $\theta_{\mathrm{UD}}$ & night      & $t_{\rm obs}$\\
HD number  &      (mas)          &            &    (UTC) \\ \hline
 90393     & $0.77 \pm 0.01$     & 2005-02-26 & 04:38:04           \\
\hline
\end{tabular}
\begin{tabular}{c}
\\
\textbf{MIDI}  ($7.9\leq \lambda \leq 13.5\micron$) \\
2 Unit Telescopes
\end{tabular}
\begin{tabular}{c c c c c}\hline
Calibrator & $\theta_{\mathrm{UD}}$ & night      & $t_{\rm obs}$\\
HD number  &      (mas)          &            &    (UTC) \\ \hline
 37160     & $2.08\pm0.20$       & 2004-12-29 & 04:12:26           \\
           &             &                    & 05:29:32           \\ \hline
 50778     & $3.95\pm0.22$       & 2004-12-29 & 06:13:08           \\
           &                     & 2004-12-31 & 02:15:59           \\
           &                     &            & 03:04:33           \\ \hline
 94510     & $2.16\pm0.11$       & 2004-12-29 & 07:47:21           \\
           &                     & 2004-12-31 & 06:31:19           \\
           &                     &            & 07:41:22           \\ \hline
107446     & $4.54\pm0.23$       & 2004-12-31 & 07:19:17
\\ \hline
\end{tabular}
\end{center}
\end{table}

%

\section{Photometric observations}
\label{photometry}

%
\begin{table*}
\caption[]{Photometry of CPD$-57\degr\,2874$ obtained at SAAO.}
\label{ta:photometry_SAAO}
\begin{center}
\begin{tabular}{cccccccccc}
\hline\noalign{\smallskip}
 JD      &$V$  & $U-B$& $B-V$  & $V-R{_c}$ & $V-I{_c}$ & $J$ & $H$ & $K$ & $L$\\
2450000+&     &      &      &       &       \\
\noalign{\smallskip}\hline\noalign{\smallskip}
  615.29&     &      &      &       &      &5.76 & 4.87 & 4.02 & 2.73 \\
  640.21&     &      &      &       &      &5.80 & 4.88 & 4.03 & 2.74 \\
  640.23&10.08& 0.30 & 1.65 & 1.23  & 2.39 &     &      &      &      \\
  811.57&     &      &      &       &      &5.74 & 4.86 & 4.02 & 2.73 \\
\noalign{\smallskip}\hline
\end{tabular}
\end{center}
\end{table*}
%


In addition to the VLTI data we also present here new broadband
photometric observations of CPD$-57\degr\,2874$. Optical
($UBVR_{c}I_{c}$) and near-IR ($JHKL$) photometric observations
were obtained quasi-simultaneously on 1997 July 10 at the
South-African Astronomical Observatory (SAAO). Additional near-IR
observations were obtained on 1997 June 15 and December 28. The
0.75--meter telescope with a single-element InSb photometer
(Carter 1990) was used in the near-IR, while the 0.5--meter
telescope with a GaAs photometer (Menzies et al. 1991) was used in
the optical region.

The data are presented in Table \ref{ta:photometry_SAAO}. {The
errors in the tabulated magnitudes and colors are not greater than
0.02 mag.} A number of standard stars were observed during the
same nights for calibration.

Our photometric results are very close to a few published
observations of the star. Drilling (1991) obtained 3 $UBV$
observations in 1972--1976 ($V$=10.20:, $B-V$=1.66, $U-B$=0.41
mag; the colon indicates either a variability suspicion or an
uncertainty of over 0.08 mag), and McGregor et al. (1988) obtained
near-IR observations on 1983 May 15 ($J=5.77$, $H=4.99$, $K=4.02$
mag). The 2MASS data obtained on 2000 January 18 (Cutri et al.
2003) are very similar ($J=5.76$, $H=4.96$, $K=4.3\pm0.3$ mag).
However, the near-IR fluxes from Swings \& Allen (1972) are very
different: $K=5.44$, $H-K=0.33$, $K-L \sim0.7$ mag. Also,
Wackerling (1970) quotes $m_{\rm vis}=8.6$ mag and $m_{\rm
pg}=10.2$ mag\footnote{m$_{pg}$ means photographic magnitude,
which is usually considered a rough analog of a B-band
photometry.}. This information is not sufficient to conclude
whether any brightness changes occurred in the early 1970s, but it
indicates that the optical and near-IR fluxes have been stable for
the last 30 years.

\begin{figure*}[t]
 \centering
  \includegraphics*[width=12.5cm,draft=false]{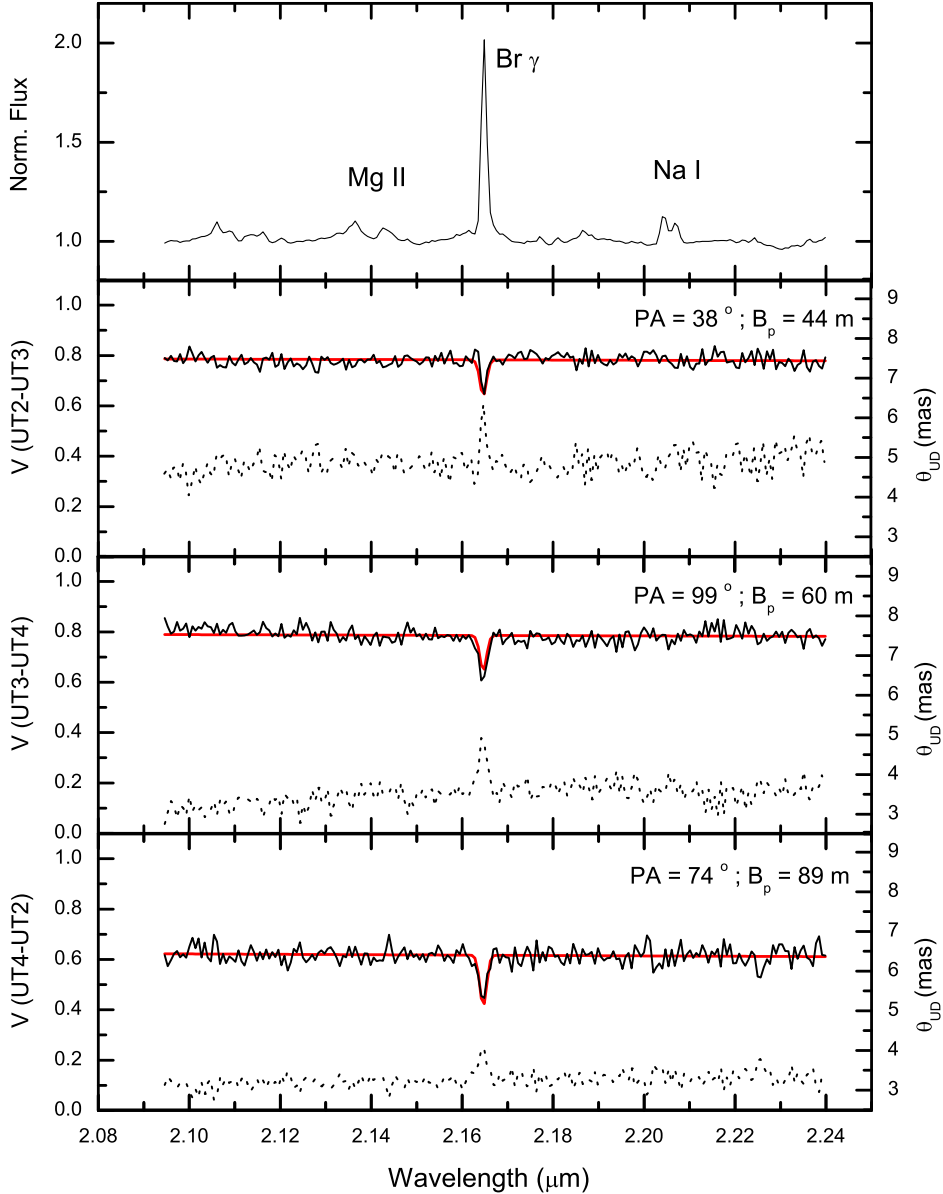}
  \caption{VLTI/AMBER observations of CPD$-57\degr\,2874$ obtained
around Br$\gamma$ with spectral resolution $R=1500$. The
normalized flux is shown in the top panel and the visibilities $V$
for each baseline (30\% best frames) are given in the other panels
(the corresponding projected baselines $B_\mathrm{p}$ and position
angles PA are indicated). The errors in $V$ are $\simeq \pm 5\%$.
The dotted lines are the uniform disk angular diameters
$\theta_\mathrm{UD}$ (to be read from the scales on the right
axis), computed from $V$ at each $\lambda$ as a zero-order size
estimate. The visibilities obtained from the elliptical Gaussian
model fit the observations quite well (smooth solid lines;
Eqs.~\ref{eq:gaussian_ellipse_model_V} and \ref{eq:model_size},
and Table~\ref{ta:models}). In contrast to the Br$\gamma$ line,
the Mg~{\sc ii} and Na~{\sc i} lines do not show any clear
signature on the visibilities.}
  \label{fig:AMBER_results}
%
\end{figure*}
%

\begin{figure*}[t]
 \centering
  \includegraphics*[width=12.5cm,draft=false]{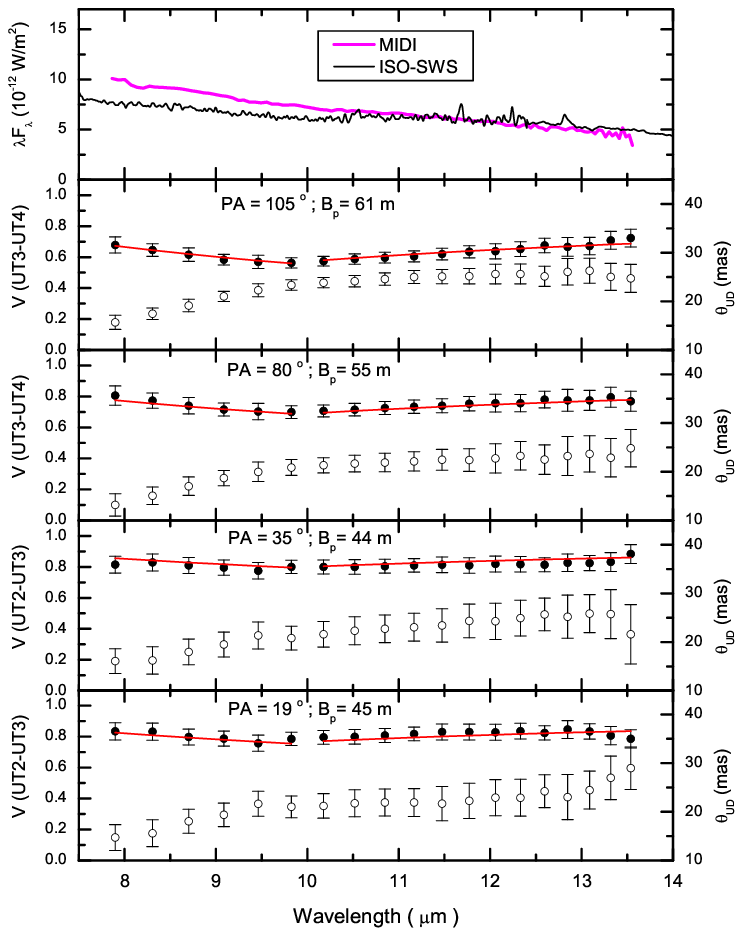}
  \caption{VLTI/MIDI observations of CPD$-57\degr\,2874$ obtained in the
mid-IR with spectral resolution $R=30$. This figure is organized
as Fig.~\ref{fig:AMBER_results}, but here $V$ and
$\theta_\mathrm{UD}$ are shown as filled and open circles,
respectively. The MIDI and ISO-SWS (Sloan et al. 2003) spectra
(top panel) do not show any clear evidence of a silicate feature
around $10\micron$. The MIDI visibilities are well fitted with an
elliptical Gaussian model (solid lines;
Eqs.~\ref{eq:gaussian_ellipse_model_V} and \ref{eq:model_size},
and Table~\ref{ta:models}).}


  \label{fig:MIDI_results}
%
\end{figure*}

{Analysis of the available photometric and spectroscopic data for
the object and its neighborhood allows us to put some constraints
on the basic parameters of the underlying star and the distance
toward it, an issue that has never been carefully addressed. The
observed set of emission lines in the optical region (H {\sc i},
He {\sc i}, Fe {\sc ii}; Zickgraf 2003, McGregor et al. 1988,
Carlson \& Henize 1979, and others) suggests that the star has an
early B spectral type, which in combination with the large optical
colour-indices implies a high reddening (see
Table~\ref{ta:photometry_SAAO}). However, the presence of a
significant amount of gas and dust in the object's CSE makes
uncertain whether the entire reddening is interstellar. On the
other hand, this is most likely the case, because the observed
$U-B$ and $B-V$ colour-indices are in agreement with the
interstellar reddening slope for the stars in the object's
direction (E$(U-B)$/E$(B-V)=0.74\pm0.06$). If we ignore the
possible impact of the CSE gas on the object's SED, then
dereddening with the above colour-index ratio gives E$(B-V)=1.85$
mag and the spectral type B$1\pm1$ (also in agreement with the
spectral line content). Moreover, strong diffuse interstellar
bands (at $\lambda5780$~\AA\ and $\lambda5797$~\AA) are present in
the spectrum, and their strengths are consistent with the E$(B-V)$
(Herbig~1993).

Adopting the typical galactic total-to-selective interstellar
extinction ratio A$_V$/E$(B-V)=3.1$ for early-type stars, we get
A$_V=5.8$ mag and the intrinsic visual brightness $V_0=4.3$ mag.
Such a brightness, in combination with the high reddening, implies
a high stellar luminosity. Since a few nearby A-type stars of
9--10 mag have negligible reddenings, there is almost no
interstellar extinction in the object's direction closer than
$\sim 1$~kpc. Even at such a distance, CPD$-57\degr$2874 would be
a supergiant ($\log L/L_{\sun} \sim 5$). An upper limit for the
distance ($\sim3$~kpc) is set by the Humphreys-Davidson luminosity
limit ($\log L/L_{\sun} \sim 6$, Humphreys \& Davidson 1979).
Thus, the most probable range for the object's distance is
$2.5\pm0.5$~kpc. It is difficult to constrain it better due to the
unknown contribution of the CSE gas to the star's brightness,
possible anomalous extinction by the CSE dust, and the absence of
high-resolution spectroscopic data that show photospheric lines
and allow us to measure their radial velocities. The distance
determination using galactic kinematic models and available radial
velocities of the emission lines is uncertain, because the line
profiles are double-peaked. The interstellar extinction law in the
object's direction indicates a patchy structure of dust in the
line of sight and hampers further improvement of the above
distance estimate.

Summarizing the above discussion, we adopt the following
parameters for CPD$-57\degr$2874: $d=2.5\pm0.5$~kpc, T$_{\rm
eff}=20000\pm3000$~K, A$_V=5.8$~mag. They lead to an estimate for
the star's radius of $R_\star=60\pm15\rsun$.

}

\section{Results}
\label{results}

Figures~\ref{fig:AMBER_results} and \ref{fig:MIDI_results} show
the spectra and visibilities obtained with AMBER and MIDI,
respectively. CPD$-57\degr\,2874$ is resolved in both spectral
regions at all projected baselines $B_\mathrm{p}$ and position
angles PA. As a zero-order size estimate these figures also show
the uniform disk angular diameters $\theta_\mathrm{UD}$ obtained
from the visibilities at each spectral channel. The size of the
region emitting the Br$\gamma$ line is larger than the region
emitting the near-IR continuum. Moreover, the mid-IR sizes are
much larger than those in the near-IR.

\begin{figure}[t]
 \centering
  \includegraphics*[width=\hsize,draft=false]{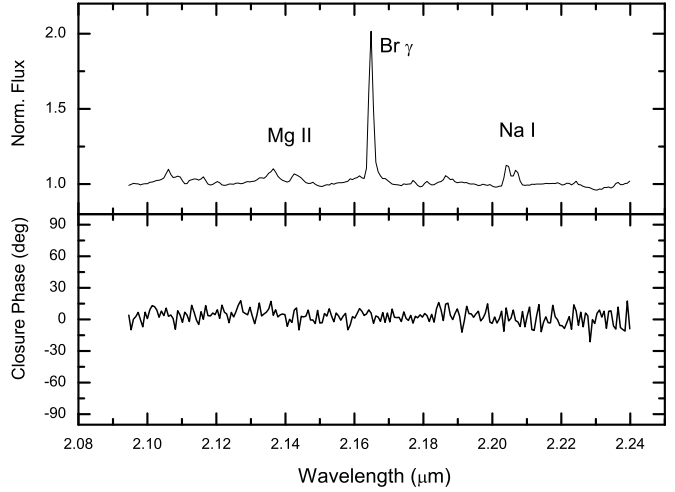}
  \caption{VLTI/AMBER closure phase for CPD$-57\degr\,2874$ obtained
around Br$\gamma$ with spectral resolution $R=1500$. Within the
noise level ($\simeq \pm 10\degr$) the closure phase is zero both
in the continuum and in the Br$\gamma$ line. This strongly
suggests that the CSE has an approximately centrally-symmetric
projected intensity distribution in the near-IR. }
  \label{fig:closure_phase}
%
\end{figure}

The AMBER observations also reveal a zero closure phase
(Fig.~\ref{fig:closure_phase}) at all wavelengths (within the
noise level of a few degrees). This is a strong indication that
the near-IR emitting regions (continuum and Br$\gamma$ line) have
an approximately centrally-symmetric intensity distribution.

Since sgB[e] stars are thought to have non-spherical winds, we
expect an elongated shape for their CSE projected onto the sky,
unless the star is seen close to pole-on. Hereafter, we show that
both AMBER and MIDI observations can indeed be well reproduced by
an elliptical Gaussian model for the CSE intensity distribution,
corresponding to visibilities of the form:
\begin{equation}\label{eq:gaussian_ellipse_model_V}
    V(u,v)=   \exp{ \left\{ \frac{-\pi^2(2a)^2}{4\ln2}[u^2+(Dv)^2] \right\} }
\end{equation}
where $u$ and $v$ are the spatial-frequency coordinates, $2a$ is
the major-axis FWHM of the intensity distribution (image plane),
and $D$ is the ratio between the minor and major axes FWHM
($D=2b/2a$). Since, in general, $2a$ forms an angle $\alpha$ with
the North direction (towards the East), $u$ and $v$ should be
replaced in Eq.~\ref{eq:gaussian_ellipse_model_V} by
($u\sin\alpha+v\cos\alpha$) and ($u\cos\alpha-v\sin\alpha$),
respectively. A preliminary analysis of $V$ at each individual
wavelength $\lambda$ showed that $D$ and $\alpha$ can be
considered independent on $\lambda$ within a given spectral band
($K$ or $N$). On the other hand, the CSE size varies with
$\lambda$, as seen from the $\theta_\mathrm{UD}$ curves in
Figs.~\ref{fig:AMBER_results} and \ref{fig:MIDI_results}.



\subsection{Size and geometry in the $K$ band}

We interpret the AMBER observations in terms of an elliptical
Gaussian model (Eq.~\ref{eq:gaussian_ellipse_model_V}) with a
chromatic variation of the size. The $\theta_\mathrm{UD}(\lambda)$
curves in Fig.~\ref{fig:AMBER_results} suggest a linear increase
of the size within this part of the $K$ band. In addition, the
AMBER visibilities decrease significantly inside Br$\gamma$,
indicating that the line-forming region is more extended than the
region responsible for the underlying continuum. Based on these
considerations, we adopted the following expression for the
major-axis FWHM:
\begin{equation}\label{eq:model_size}
    2a(\lambda)=2a_0 + C_1(\lambda-\lambda_0) +
C_2 \exp{ \!\!\left[
{\!-4\ln2\left(\frac{\lambda-\lambda_{\mathrm{Br}\gamma}}{\Delta
\lambda}\right)^2} \right] }
\end{equation}
where $2a_0$ is the major-axis FWHM at a chosen reference
wavelength $\lambda_0(=2.2\micron)$, and $C_1$ is the slope of
$2a(\lambda)$. The size-increase within Br$\gamma$ is modeled by a
Gaussian with an amplitude $C_2$ and FWHM $\Delta\lambda$,
centered at $\lambda_{\mathrm{Br}\gamma}=2.165\micron$.
Figure~\ref{fig:AMBER_results} shows a rather good fit of this
model to the observed visibilities in both the continuum and
inside Br$\gamma$. The parameters derived from the fit are listed
in Table~\ref{ta:models}.


%

\subsection{Size and geometry in the $N$ band}

Similar to the analysis of the AMBER visibilities, we interpret
the MIDI observations of CPD$-57\degr\,2874$ in terms of an
elliptical Gaussian model (Eq.~\ref{eq:gaussian_ellipse_model_V})
with a size varying linearly with $\lambda$ as given in
Eq.~\ref{eq:model_size} (for the analysis of the MIDI data the
parameter $C_2$ is set to zero). Additionally, since the
$\theta_\mathrm{UD}$ curves show a stronger $\lambda$-dependence
between 7.9 and 9.8 $\micron$ compared to the region between 10.2
and 13.5 $\micron$ (see Fig.~\ref{fig:MIDI_results}), we performed
an independent fit for each of these two spectral regions. The
adopted elliptical Gaussian model also provides a good fit to the
MIDI visibilities as shown in Fig.~\ref{fig:MIDI_results}. The
parameters corresponding to the fit in the two spectral regions
within the $N$ band are listed in Table~\ref{ta:models}.

\smallskip

\begin{figure*}[ht]
 \centering
  \includegraphics*[width=150mm]{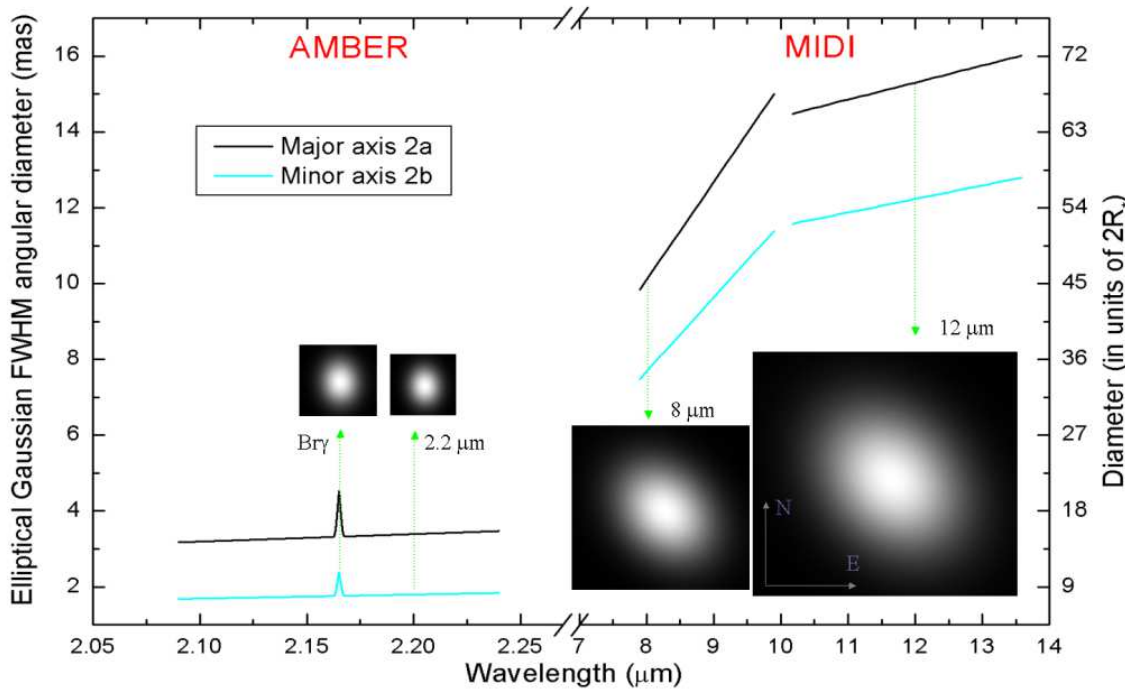}
  \caption{This figure illustrates the results (size and orientation)
derived from the fit of an elliptical Gaussian model
(Eqs.~\ref{eq:gaussian_ellipse_model_V} and \ref{eq:model_size})
to the VLTI/AMBER and VLTI/MIDI visibilities (the corresponding
fit parameters are listed in Table~\ref{ta:models}). The scale in
the right is given in stellar diameters ($2R_*$), where the radius
$R_*$ is estimated to be $60\pm15\rsun$ (Sect.~\ref{photometry}).}
  \label{fig:sizes_AMBER_MIDI_ellipses}
\end{figure*}

{To illustrate our results the model parameters given in
Table~\ref{ta:models} are visualized in
Fig.~\ref{fig:sizes_AMBER_MIDI_ellipses}.}

\begin{table*}[ht]
 \centering
 \begin{minipage}[]{\hsize}
   \renewcommand{\footnoterule}{}
 \centering
\caption[]{Model parameters and $\chi^2_\mathrm{red}$ (reduced
chi-squared) derived from the fit of an elliptical Gaussian
(Eqs.~\ref{eq:gaussian_ellipse_model_V} and \ref{eq:model_size})
to the VLTI/AMBER and VLTI/MIDI visibilities. Angular sizes (in
mas) correspond to FWHM diameters. The errors of the fit
parameters include the calibration errors of the instrumental
transfer function ($\simeq5\%$), estimated from all calibrator
stars observed. }
\label{ta:models} \vspace{0.1cm}
\begin{tabular}{lccccccccc}
\hline\noalign{\smallskip}


Instrument & $\lambda_0$ & major axis       & $C_1$        & position        &  $D=2b/2a$ & minor axis

& \textbf{Br$\gamma$:} $C_2$

 & \textbf{Br$\gamma$:} $\Delta\lambda$& $\chi^2_\mathrm{red}$\\
           & ($\mu$m)    &   $2\,a_0$ (mas) &  (mas/$\mu$m)&  angle $\alpha$ &            & $2\,b_0$ (mas) &

(mas)\footnote{Corresponding to a FWHM major axis $2a=4.5\pm0.3$
mas and minor axis $2b=D*2a=2.4\pm0.1$ mas at the center of the
Br$\gamma$ line ($\lambda_{\mathrm{Br}\gamma}=2.165\micron$). }& ($10^{-3}\mu$m)       &         \\


\noalign{\smallskip}\hline\noalign{\smallskip} $\!\!\!$AMBER &
$\!\!\!$ 2.2 & $\!\!\!$ $3.4\pm0.2$ & $\!\!\!$ 1.99$\pm$0.24&
$\!\!\!$ $173\degr\pm9\degr$ &
$\!\!\!$ 0.53$\pm$0.03 & $\!\!\!$ 1.8$\pm$0.1 & $\!\!\!$ 1.2$\pm$0.1 & $\!\!\!$  1.8$\pm$0.2 & 0.7\\
$\!\!\!$MIDI ($<10\micron$)  & $\!\!\!$ 8.0 & $\!\!\!$10.1$\pm$0.7
& $\!\!\!$ 2.58$\pm$0.41& $\!\!\!$ $145\degr\pm6\degr$ &
$\!\!\!$ 0.76$\pm$0.11 & $\!\!\!$ 7.7$\pm$1.0 &  & & 0.1\\
$\!\!\!$MIDI ($>10\micron$)  & $\!\!\!$ 12.0& $\!\!\!$15.3$\pm$0.7
& $\!\!\!$ 0.45$\pm$0.22& $\!\!\!$ $143\degr\pm6\degr$ &
$\!\!\!$ 0.80$\pm$0.10 & $\!\!\!$ 12.2$\pm$1.1$\,\,\,$ &  & & 0.1\\
\noalign{\smallskip}\hline\noalign{\smallskip}
\end{tabular}
\end{minipage}
\end{table*}
%


%
%

\section{Discussion and conclusions}
\label{discussion}

{Our analysis of the VLTI spectro-interferometric data presented
in Sect.~\ref{results} supports the hypothesis of a non-spherical
CSE for sgB[e] stars.}

In particular, the MIDI observations suggest that the hot-dust
emission originates in an elongated structure (probably in an
equatorial disk), which is in agreement with previous polarization
measurements from Yudin \& Evans (1998). After correction for the
interstellar polarization, Yudin (private communication) estimated
an intrinsic polarization position angle $\simeq45\degr-55\degr$.
Interestingly, within the error bars this angle is perpendicular
to the major-axis PA we derived from the MIDI data
($\alpha\simeq144\degr$; see Table~\ref{ta:models}), as is
expected from a disk-like dusty CSE. Under the disk hypothesis,
the measured mid-IR flattening ($D\simeq0.76-0.80$; see
Table~\ref{ta:models}) allows us to estimate an intermediate
viewing angle for the non-spherical CSE ($i\sim30\degr-60\degr$).

The contemporaneous recording of the AMBER and MIDI data enables
us to compare the CSE structure in the near- and mid-IR. As shown
in Table~\ref{ta:models}, the size, flattening, and orientation of
the elliptical Gaussian model significantly changes from the $K$
to the $N$ band. For example, the region emitting the mid-IR flux
($2a\ge10~\mathrm{mas} = 25~\mathrm{AU}$\footnote{Adopting a
distance $d=2.5$~kpc; see Sect.\ref{photometry}.} at
$\lambda\ge8\micron$) is more than 2.5 times larger than the one
emitting the near-IR continuum flux ($2a\simeq 3.4~\mathrm{mas} =
8.5~\mathrm{AU}$ at $\lambda \simeq 2.2\micron$).

If we correct the influence of the continuum on the visibility
measured in Br$\gamma$ (Malbet et al. 2005), we estimate the size
(minor $\times$ major axes) of the region responsible for the pure
Br$\gamma$ emission to be $\simeq2.8\times5.2$~mas (or
$\simeq7.0\times13.0$~AU). This size is $\simeq55\%$ larger than
that of the underlying near-IR continuum, but more than $2$ times
smaller than the mid-IR emitting region ($\lambda\ge8\micron$).
Near-IR diameters of $\sim10$~AU correspond to $\sim36R_\star$
(assuming $R_\star = 60 \rsun$; see Sect.\ref{photometry}). This
measurements are compatible the theoretical CSE diameters computed
by Stee \& Bittar (2001) for a classical Be star, though our data
show a larger difference between the Br$\gamma$ and continuum
sizes.

The differences in flattening and position angle of the elliptical
models fitted to the AMBER and MIDI data are in agreement with the
two-component CSE paradigm suggested for sgB[e] stars (Zickgraf et
al. 1985). The mid-IR flux is probably solely due to dust emission
from an equatorial disk. By contrast, the near-IR continuum flux
distribution probably results from a complex interplay among the
radiation from the central star, the tail of hot-dust emission
($T_\mathrm{dust} \simeq 1500$ K), and the free-free and
free-bound emission from the fast polar wind and the disk-wind
interaction. The Br$\gamma$ emission does not necessarily follow
the same geometry.

{A detailed investigation of the CSE geometry in the near-IR
(continuum and Br$\gamma$) requires additional interferometric
observations covering a larger range of baselines and position
angles. In addition, we believe that further MIDI observations at
baselines longer than $\simeq80$~m should be performed to obtain
higher spatial resolution of the innermost parts of the dusty CSE.
This would allow one to investigate more deeply how close to the
hot central star (T$_{\rm eff}\simeq20000$~K) the dust is formed.}

{Moreover, the combination of interferometric resolution,
multi-spectral wavelength coverage and relatively high spectral
resolution now available from the VLTI, requires de development of
interferometry-oriented and physically-consistent models for
sgB[e] stars. We hope that the present work will motivate the
development of such models, as well as open the door for new
spectro-interferometric observations of these complex and
intriguing objects.}

\section{Acknowledgments}
A.D.S.\ acknowledges the Max-Planck-Institut f\"{u}r
Radio\-astro\-nomie for a postdoctoral fellowship. We thank the
JMMC-France user-support for information about the ASPRO software.
We are indebted to Dr.\ R.\ V.\ Yudin for his calculations on the
intrinsic polarization vector.



{}

\end{document}